\documentclass[twocolumn,showpacs,showkeys,preprintnumbers,amsmath,amssymb]{revtex4}

\usepackage{graphicx}
\usepackage{dcolumn}
\usepackage{bm}

\begin{document}

\preprint{}

\title{Reversed propagation dynamics of Laguerre-Gaussian beams in left-handed materials}
\author{Hailu Luo$^{1,2}$}\email{hailuluo@gmail.com}
\author{Zhongzhou Ren$^{1}$}
\author{Weixing Shu$^{2}$}
\author{Shuangchun Wen$^{2}$}
\affiliation{$^{1}$ Department of Physics, Nanjing University,
Nanjing 210008, China\\
$^{2}$ School of Computer and Communication, Hunan University,
Changsha 410082, China}
\date{\today}

\begin{abstract}
On the basis of angular spectrum representation, the reversed
propagation dynamics of Laguerre-Gaussian beam in left-handed
materials (LHMs) is presented. We show that negative phase velocity
gives rise to a reversed screw of wave-front, and ultimately leads
to a reversed rotation of optical vortex. Furthermore, negative
Gouy-phase shift causes an inverse spiral of Poynting vector. It is
found that the Laguerre-Gaussian beam in LHMs will present the same
propagation characteristics as the counterpart with opposite
topological charges in regular right-handed materials (RHMs). The
momentum conservation theorem insures that the tangential component
of the wave momentum at the RHM-LHM boundary is conserved. It is
shown that although the linear momentum reverses its direction, the
angular momentum remains unchanged.
\end{abstract}

\maketitle

\section{Introduction}\label{Introduction}
Almost 40 years ago, Russian scientist Victor Veselago proposed that
a material with electric permittivity $\varepsilon<0$ and magnetic
permeability $\mu<0$, would reverse all known optical
properties~\cite{Veselago1968}. He termed these media as left-handed
materials (LHMs) since the wave vector ${\bf k}$, forms a
left-handed triplet with the vectors ${\bf E}$ and ${\bf H}$. That
is, phase velocity and the Poynting vector are antiparallel, which
consequently results in counter-intuitive phenomena such as
reversals of the conventional Doppler shift and Cherenkov radiation
as well as reversed refraction. Veselago pointed out that
electromagnetic waves incident on a planar interface between a
regular right-handed material(RHM) and a LHM will undergo negative
refraction. Hence a LHM planar slab can act as a lens and focus
waves from a point source. Recently, Pendry extended Veslago's
analysis and further predicted that a LHM slab can amplify
evanescent waves and thus behaves like a perfect
lens~\cite{Pendry2000}. Pendry proposed that the amplitudes of
evanescent waves from a near-field object could be restored at its
image. Therefore, the spatial resolution of the superlens can
overcome the diffraction limit of conventional imaging systems and
reach the sub-wave length scale. The physical realization of such a
LHM was demonstrated only recently for a novel class of engineered
composite
materials~\cite{Smith2000,Shelby2001,Parazzoli2003,Houck2003}.

After the first experimental observation of negative refraction,
intriguing and counterintuitive phenomenon in LHMs,  such as
amplification of evanescent waves~\cite{Pendry2000,Fang2005},
unusual photon tunneling~\cite{Zhang2002,Kim2004}, and negative
Goos-H\"{a}nchen shift~\cite{Kong2002,Berman2002} have attracted
much attention. Here we want to explore the reversed propagation
dynamics of Laguerre-Gaussian beams in LHMs. The propagation of
Laguerre-Gaussian beam has been investigated in conventional
RHMs~\cite{Basistiy1995,Rozas1997,Curtis2003,Grier2003}. Such beams
have a phase dislocation on the beam axis that in related literature
is sometimes referred to as an optical vortex~\cite{Curtis2003}. For
a general Laguerre-Gaussian beam the Poynting vector has an
azimuthal component. This means that there is an energy flow along
the circumference of the beam as it propagates, giving rise to an
orbital angular momentum~\cite{Allen1992}. It is found that the
spiral of the Poynting vector of a Laguerre-Gaussian beam is
proportional to the Gouy-phase shift~\cite{Padgett1995,Allen2000}.
It is known that an electromagnetic beam propagating through a focus
experiences an additional $\pi$ phase shift with respect to a plane
wave. This phase anomaly was discovered by Gouy in 1890 and has
since been referred to as the Gouy-phase shift~\cite{Siegman1986}.
Because of the negative index, however, we can expect an reversed
Gouy-phase shift in LHMs~\cite{Luo2007b}. Hence it will be
interesting for us to describe in detail how the Poynting vector
evolves as it propagates and how the reversed Gouy-phase shift
affects its spiral in LHMs.

In this work, we will reveal reversed propagation dynamics of
Laguerre-Gaussian beam in LHMs, such as inverse screw of wave-front,
inverse spiral of Poynting vector, and inverse rotation of vortex
field. First, starting from the representation of plane-wave angular
spectrum, we obtain the analytical description for a
Laguerre-Gaussian beam propagating in LHMs. Our formalism permits us
to introduce the reversed Gouy-phase shift to describe the wave
propagation. Next, we will recover how the wave-front and Poynting
vector evolves, and how the reversed Gouy-phase shift affects their
propagation behavior. Then, we attempt to investigate how the
negative index influences the linear momentum and angular momentum
of Laguerre-Gaussian beams. Finally, we will explore how the
negative index gives rise to the reversed rotation of the vortex
field. For a comparison, the corresponding propagation
characteristics in RHMs will also been discussed.

\section{The Paraxial propagation of a Laguerre-Gaussian beam}\label{II}
To investigate the propagation dynamics of a Laguerre-Gaussian beam
in LHMs, we use the Maxwell's equations to determine the field
distribution both inside and outside the LHM. We consider a
monochromatic electromagnetic field ${\bf E}({\bf r},t) = Re [{\bf
E}({\bf r})\exp(-i\omega t)]$ and ${\bf B}({\bf r},t) = Re [{\bf
B}({\bf r})\exp(-i\omega t)]$ of angular frequency $\omega$
propagating from the RHM to the LHM. The field can be described by
Maxwell's equations
\begin{eqnarray}
\nabla\times {\bf E} &=& - \frac{\partial {\bf B}}{\partial t},
~~~{\bf B} = \mu_0 \boldsymbol{\mu}\cdot{\bf H},\nonumber\\
\nabla\times {\bf H} &=& \frac{\partial {\bf D}}{\partial
t},~~~~~{\bf D} =\varepsilon_0 \boldsymbol{\varepsilon} \cdot {\bf
E}. \label{maxwell}
\end{eqnarray}
From the Maxwell's equations, we can easily find that the wave
propagation is only permitted in the medium with $\varepsilon,
\mu>0$ or $\varepsilon,\mu<0$. In the former case, ${\bf E}$,
${\bf H}$ and ${\bf k}$ form a right-handed triplet, while in the
latter case, ${\bf E}$, ${\bf H}$ and ${\bf k}$ form a left-handed
triplet.

We introduce the Lorentz-gauge vector potential to describe the
propagation characteristics of Laguerre-Gaussian in RHMs and LHMs.
The vector potential of the beam propagating in the $+z$ direction
can be written in the form
\begin{equation}
{\bf A}=A_0(\alpha{\bf e}_x+\beta{\bf e}_y)u_{p,l}({\bf r})\exp(i
k z-i\omega t),\label{ca}
\end{equation}
where $A_0$ is a complex amplitude, ${\bf e}_x$ and ${\bf e}_y$ are
unit vectors, $k =n_{R,L}\omega/c$, $c$ is the speed of light in
vacuum, $n_R=\sqrt{\varepsilon_R\mu_R}$ and
$n_L=-\sqrt{\varepsilon_L\mu_L}$ are the refractive index of RHM and
LHM, respectively~\cite{Veselago1968}. The coefficients  $\alpha$
and  $\beta$ satisfying
$\sigma=i(\alpha\beta^\ast-\alpha^\ast\beta)$, are the polarization
operators with $\sigma=\pm1$ for left-handed and right-handed
circularly polarized light.

When the field distribution is specified at a boundary surface or
a transverse plane, one can obtain a unique solution of the
electric field of the wave propagating in the $+z$ direction.
Here, we assume that the transverse electric field at the $z=0$
plane is given by a Laguerre-Gaussian function as follows:
\begin{eqnarray}
u(r,\varphi,0)=\frac{C_{pl}}{w_{0}}
\left[\frac{\sqrt{2}r}{w_{0}^2}\right]^{|l|} L_p^{|l|}\left[\frac{2
r^2}{w_{0}^2}\right]\exp\left[\frac{r^2}{w_{0}^2}-il\varphi\right].\label{F0}
\end{eqnarray}
A Laguerre-Gaussian beam has two mode indices to fully describe the
mode: $l$ and $p$. A given mode will have $l$ complete cycles of
phase $2\pi$ upon going around the mode circumference, so that $l$
is known as the azimuthal index. The index $p$ gives the number
$p+1$ of radial nodes. Laguerre-Gaussian light beams are well known
to possess orbital angular momentum due to an $\exp [il\varphi]$
phase term, where $\varphi$ is the azimuthal phase. This obital
angular momentum $l \hbar$ is distinct from the spin angular
momentum due to the polarization state of the
light~\cite{Allen1992}.

From the point of view of Fourier optics, we know that if the
Fourier component at the $z=0$ plane represents the angular spectrum
that the transverse component of the wave propagating in the half
space $z>0$ should have. Then, the field in the region $z>0$ can be
expressed by an integral of the plane wave components associated
with the angular spectrum given at the $z>0$
plane~\cite{Goodman1996}. The angular spectrum is related to the
boundary distribution of the field by means of the relation
\begin{eqnarray}
\tilde{{u}}(k)=\int_0^\infty d r r J_l(k r)u(
r,\varphi,0),\label{as}
\end{eqnarray}
where $J_l$ is the first kind of Bessel function with order $l$. The
two-dimensional Fourier transformations of Eq.(\ref{as}) can be
easily obtained from an integration table~\cite{Gradshteyn1980}. In
fact, after the field on the plane $z=0$ is known, Eq.~(\ref{F0})
together with Eq.~(\ref{as}) provides the expression of the field in
the space $z>0$, which yields
\begin{equation}
{u}(r,\varphi,z )=\int_0^\infty d k k \exp \bigg(-\frac{i k^2
z}{2n_{R,L} k_0}\bigg)J_l(kr)\tilde{u}(k).\label{field}
\end{equation}
which is a standard two-dimensional Fourier
transform~\cite{Goodman1996}. The field $u({\bf r}, z)$ is the
slowly varying envelope amplitude which satisfies the paraxial
wave equation
\begin{equation}
\bigg[i\frac{\partial}{\partial z}+\frac{1}{2 n_{R,L}
k_0}\nabla_\perp^2 \bigg]
 u({\bf r},z)=0,\label{pe}
\end{equation}
where $\nabla_\perp=\partial_x {\bf e}_x+ \partial_y {\bf e}_y$.
From Eq.~(\ref{pe}) we can find that the field of paraxial beam in
LHMs can be written in the similar way to that in RHMs, while the
sign of the refractive index is negative.

The gauge condition on the vector and scalar potentials takes the
form $\phi=(i/k)\nabla\cdot {\bf A}$. The electric and magnetic
fields are obtained from the potentials as
\begin{eqnarray}
{\bf E}({\bf r},t)&=&-\frac{\partial{\bf A}}{\partial t}-\nabla
\phi=A_0\bigg[i \omega (\alpha{\bf e}_x+\beta{\bf e}_y)u\nonumber\\
&&-\bigg(\alpha \frac{\partial u}{\partial x}+\beta \frac{\partial
u}{\partial y}\bigg){\bf e}_z\bigg]\exp(ikz-i\omega t)
,\label{Efield}
\end{eqnarray}
\begin{eqnarray}
{\bf B}({\bf r},t)&=&\nabla \times{\bf A}
=A_0\bigg[-i k (\beta{\bf e}_x-\alpha{\bf e}_y)u\nonumber\\
&&+\bigg(\beta \frac{\partial u}{\partial x}-\alpha \frac{\partial
u}{\partial y}\bigg){\bf e}_z\bigg]\exp(ikz-i\omega t)
.\label{Hfield}
\end{eqnarray}
These field expressions neglect terms in each component that are
smaller than those retained in accordance with the paraxial
approximation~\cite{Lax1975}. The $z$ components are smaller than
the $x$ and $y$ components by a factor of order $1/k w_0$. It is
readily verified that the fields satisfy Maxwell's equations. Note
that the Cartesian derivatives can be converted to polar $r$ and
$\varphi$ derivatives in the usual way.

To be uniform throughout the following analysis, we introduce
different coordinate transformations $z_i^\ast (i=1,2)$ in the RHM
and the LHM, respectively. First we want to explore the field in the
RHM. Without any loss of generality, we assume that the input waist
locates at the object plane $z=-a$ and $z_1^\ast=z+a$. The field in
the RHM can be written as
\begin{eqnarray}
u_{pl}^R=&&\frac{C_{pl}}{w(z_1^\ast)}
\left[\frac{\sqrt{2}r}{w^2(z_1^\ast)}\right]^{|l|}
L_p^{|l|}\left[\frac{\sqrt{2}r}{w^2(z_1^\ast)}\right]
\exp\bigg[\frac{-r^2}{w^2(z_1^\ast)}\bigg]\nonumber\\
&&\times \exp\bigg[i n_R k_0 z_1^\ast+\frac{-i n_R k_0 r^2
z_1^\ast}{R(z_1^\ast)}\bigg]\exp[-il\varphi]\nonumber\\ &&\times
\exp[-i (2p+|l|+1)\arctan (z_1^\ast/z_R)],\label{F1}
\end{eqnarray}
\begin{eqnarray}
w(z_{1}^\ast)=w_0\sqrt{1+(z_{1}^\ast/z_R)^2},~~R(z_{1}^\ast)=z_1^\ast+\frac{z_R^2}{z_1^\ast}.
\end{eqnarray}
Here $C_{pl}$ is the normalization constant, $L_p^l[2
r^2/w_{1}^2(z_1^\ast)]$ is a generalized Laguerre polynomial, $z_R=
n_R k_0 w_0^2 /2$ is the Rayleigh length, $w(z_{1}^\ast)$ is the
beam size and $R(z_{1}^\ast)$ the radius of curvature of the wave
front. The last term in Eq.~(\ref{F1}) denotes the Gouy phase which
is given by $\Phi_1=-(2p+|l|+1)\arctan(z_1^\ast/z_R)$.

We are now in a position to calculate the field in LHM. In fact, the
field in the RHM-LHM boundary can be easily obtained from
Eq.~(\ref{F1}) by choosing $z=0$. The plane-wave spectrum of the
Laguerre-Gaussian beam can be obtained by performing the
two-dimensional Fourier transform in Eq.~(\ref{as}). After the
plane-wave spectrum on the plane $z=0$ is known, Eq.~(\ref{field})
provides the expression of the field in the space $z>0$. For
simplicity, we assume that the wave propagates through the boundary
without reflection, the field in the LHM can be written as
\begin{eqnarray}
u_{pl}^L=&&\frac{C_{pl}}{w(z_2^\ast)}
\left[\frac{\sqrt{2}r}{w^2(z_2^\ast)}\right]^{|l|}
L_p^{|l|}\left[\frac{\sqrt{2}r}{w^2(z_2^\ast)}\right]
\exp\bigg[\frac{-r^2}{w^2(z_2^\ast)}\bigg]\nonumber\\
&&\times \exp\left[i n_L k_0 z_2^\ast+\frac{-i n_L k_0 r^2
}{R(z_2^\ast)}\right]\exp[-il\varphi]\nonumber\\&&\times \exp[-i
(2p+|l|+1)\arctan (z_2^\ast/z_L)],\label{F2}
\end{eqnarray}
\begin{eqnarray}
w(z_{2}^\ast)=w_0
\sqrt{1+(z_{2}^\ast/z_L)^2},~~R(z_{2}^\ast)=z_2^\ast+\frac{z_L^2}{z_2^\ast}.\label{w2}
\end{eqnarray}
Here $z_2^\ast=z-(1-n_L/n_R)a$ and $z_L= n_L k_0 w_0^2 /2$ is the
Rayleigh length in LHM. The beam size $w(z_2^\ast)$ and the radius
of curvature $R(z_2^\ast)$ are given by Eq.~(\ref{w2}). The
Gouy-phase shift in LHM is given by $\Phi_G=-(2p+|l|+1)\arctan
(z_2^\ast/z_L)$. Because of the negative index, the reversed
Gouy-phase shift should be introduced. A more intuitive
interpretation of the reversed Gouy phase can be given in terms of a
geometrical quantum effect~\cite{Hariharan1996} or the uncertainty
principle~\cite{Feng2001}. As can be seen in the following section,
the inverse Gouy-phase shift will give rise to an inverse spiral of
Poynting vector.

\begin{figure}
\includegraphics[width=8cm]{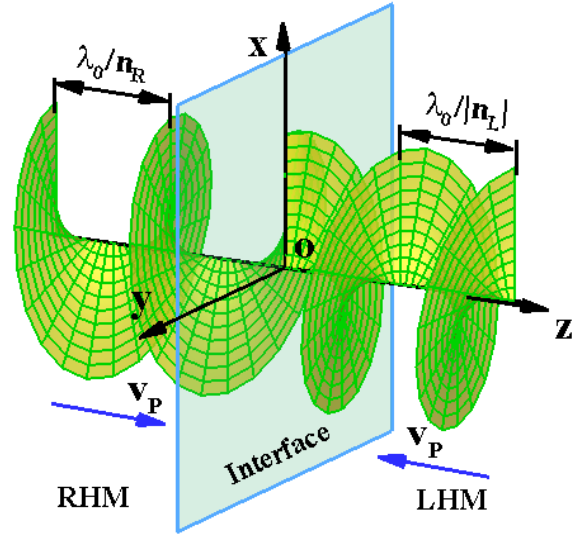}
\caption{\label{Fig1} (Color online) The helical wave front for
Laguerre-Gaussian beam with $l=1$ result from an azimuthal phase
structure of $\exp[-i\varphi]$. In the LHM, the phase velocity ${\bf
v}_p$ reverses its direction. The wave-fronts exhibit anti-clockwise
screw in the RHM, while present clockwise screw in the LHM.}
\end{figure}

For a Laguerre-Gaussian beam with $l\neq 0$, the on-axis phase form
$\exp [il\varphi]$ results in that the surfaces of wave-front have
helical form. Specifically, $l$ refers to the number of complete
cycles of phase $2\pi$ upon going around the beam circumference. Now
let us to study the screw of the wave front. Here the sense of the
positive angles is chosen as anticlockwise, while negative angles
are considered in the clockwise direction.  In the regular RHM, the
constant wavefront satisfies
\begin{equation}
n_R k_0 z_1^\ast+\frac{-i n_R k_0 r^2
}{R(z_1^\ast)}-l\varphi+\Phi_G=const.
\end{equation}
The schematic view of the wave front is a three-dimensional screw
surface of ($r\cos\varphi$, $r \sin\varphi$, $z$). The plotting
range of $r$ is from $0$ to $5w_0$ with the interval of $\Delta
r=0.5 w_0$ and that of $n_R k_0 z$ is from $-4\pi$ to $0$ with the
interval of $n_R k_0 \Delta z=0.1\pi$. The wavefront structure
exhibits a anticlockwise-screw type with a pitch of $\lambda_0/n_R$
along the $+z$ axis.

Next we explore the screwing fashion of wave front in the LHM. The
constant wavefront satisfies
\begin{equation}
n_L k_0 z_2^\ast+\frac{-i n_L k_0 r^2
}{R(z_2^\ast)}-l\varphi+\Phi_G=const.
\end{equation}
The plotting range of $r$ is from $0$ to $5w_0$ with the interval of
$\Delta r=0.5 w_0$ and that of $n_L k_0 z$ is from $0$ to $-4\pi$
with the interval of $n_L k_0 \Delta z=-0.1\pi$. The wave-front
structure is a clockwise-screw type with a pitch of
$\lambda_0/|n_L|$ along the $+z$ axis. Figure~\ref{Fig2} shows a
typical form of a helical wavefront structure from the RHM to the
LHM. At the RHM-LHM interface, the wave front will reverse its
screwing fashion. It is intriguing to observe that the wave-front of
Laguerre-Gaussian beam with $l=1$ in LHMs will exhibit the same
skewing fashion as the counterpart with $l=-1$ in RHMs. As can be
seen in the following section, the inverse screw of wavefront will
result in an inverse rotation of optical vortex.

\section{Poynting vector and angular momentum}
The propagation characteristics of electromagnetic fields are
closely linked to their local energy flow, which is usually
discussed by use of the Poynting vector. There has been considerable
interest in orbital angular momentum of Laguerre-Gaussian
beams~\cite{Allen1992} relating to Poynting vector in free space.
The Poynting vector has a magnitude of energy per second per unit
area and a direction which represents the energy flow at any point
in the field. The time average Poynting vector, ${\bf S}$ can be
written as
\begin{equation}
{\bf S}=\frac{1}{2}\text{Re}[{\bf E}\times{\bf H}^\ast].\label{PV}
\end{equation}
The spiral of the Poynting vector in free space or regular RHMs has
been discussed
extensively~\cite{Allen1992,Padgett1995,Allen2000,Volyar1999,London2003,Padgett2003}.

Now a question arise: what happens in LHMs with simultaneously
negative permeability and permittivity? The potential interests
encourage us to derive a general expression to describe the Poynting
vector in RHMs and LHMs. Substituting the expression of
Eqs.~(\ref{Efield}) and (\ref{Hfield}) into Eq.~(\ref{PV}) we find
\begin{eqnarray}
S_r&=&\frac{1}{\mu\mu_0} \frac{\omega k r}{R}|u|^2,\nonumber\\
S_\varphi&=&\frac{1}{\mu\mu_0} \left[\frac{\omega
l}{r}|u|^2-\frac{1}{2}\omega \sigma\frac{\partial |u|^2}{\partial r}\right],\nonumber\\
S_z &=&\frac{1}{\mu\mu_0}\omega k |u|^2.\label{PVD}
\end{eqnarray}
Here the component $S_r$, relates to the spread of the beam as it
propagates. The azimuthal component $S_\varphi$ describes the energy
flow that circulates around the propagating axis. The presence of
this flow is due to the existence of the longitudinal components
$E_z$ and $H_z$ of the field. The first term of the azimuthal
component depends on $l$, where $l\hbar$ has been identified as the
orbital angular momentum per photon~\cite{Allen1992}. Its second
term relates to the contribution of polarization and intensity
gradient. The contribution of circular polarization will lead to a
spin angular momentum of the beam. The axial component $S_z$
describes the energy flow that propagates along the $+z$ axis.

Next, we attempt to explore the angular momentum in LHMs. Since the
dispersion cannot be ignored in a causal system with a negative
index of refraction. The momentum conservation theorem should be
derived from the Maxwell equations and the Lorentz force and is
given by~\cite{Kong2005}
\begin{equation}
\nabla \cdot {\bf T}+ \frac{\partial {\bf G}}{\partial t}=-{\bf
F}.\label{mct}
\end{equation}
where ${\bf G}$ is the momentum density vector and ${\bf F}$ is the
force density. The momentum flow ${\bf T}$ also referred as the
Maxwell stress tensor. It is well-known that a material contribution
to the energy density accompanies the propagation of electromagnetic
energy in dispersive materials. Analogously, there exists a
corresponding material contribution to the wave momentum. Thus the
momentum conservation equation for the electromagnetic wave can be
written in the form~\cite{Kemp2005}
\begin{eqnarray}
{\bf G}&=&\frac{1}{2}\text{Re}\left[\varepsilon \mu {\bf
E}\times{\bf H}^\ast+\frac{\bf k}{2}\left(\frac{\partial
\varepsilon}{\partial \omega}|{\bf E}|^2+\frac{\partial
\mu}{\partial \omega}|{\bf
H}|^2\right)\right],\nonumber\\
{\bf T}&=&\frac{1}{2}\text{Re} [({\bf D}\cdot{\bf E}^\ast+{\bf
B}\cdot{\bf H}^\ast)I
-({\bf D} {\bf E}^\ast+{\bf B} {\bf H}^\ast)],\nonumber\\
{\bf F}&=&\frac{1}{2}\text{Re}[\rho_e {\bf E}^\ast+{\bf J}\times{\bf
B}^\ast+\rho_m {\bf H}^\ast+{\bf M}\times {\bf
D}^\ast].\label{momentum}
\end{eqnarray}
Here the momentum density ${\bf G }$ contains the Minkowski momentum
${\bf G}_M={\bf D}\times {\bf B}$ plus material dispersion terms.
The tensor $I$ is $3\times3$ identity matrix. The electric and
magnetic polarization vectors are give by ${\bf P}_e=\varepsilon_0
(\varepsilon-1){\bf E}$  and ${\bf P}_m=-\mu_0 (\mu-1){\bf H}$,
respectively. Bound electric current ${\bf J}=\partial {\bf
P}_e/\partial t $ and bound electric charge $\rho_e=\nabla\cdot{\bf
P}_e$ have been accounted. Similarly, bound magnetic current ${\bf
M}=\partial {\bf P}_m/\partial t $ and bound magnetic charge
$\rho_m=\nabla\cdot{\bf P}_m$ should be introduced to describe the
angular momentum flow in LHMs.

Now we want to enquire: how to determine directions of the momentum
density and the momentum flow? It is well known that the time-domain
energy density in a frequency nondispersive medium are defined as
$W=\frac{1}{2}[{\bf D}\cdot {\bf E}+{\bf E}\cdot {\bf H}]$.
Obviously, the energy density in LHMs would be negative if the
permittivity and permeability were negative. Hence, the energy
density in a frequency dispersive medium is defined
as~\cite{Jackson1999,Landau1984}
\begin{equation}
W=\frac{1}{4}\left[\frac{\partial (\varepsilon \omega)}{\partial
\omega}|{\bf E}|^2+\frac{\partial (\mu \omega)}{\partial
\omega}|{\bf H}|^2\right].\label{energy}
\end{equation}
In principle, the energy density can be decomposed into electric and
magnetic parts. The positive electric and magnetic energy requires
$\partial (\varepsilon \omega)/\partial \omega>0$ and $\partial (\mu
\omega)/\partial \omega>0$. Subsequent calculations of
Eq.~(\ref{momentum}) show that both the momentum density ${\bf G }$
and the momentum flow ${\bf T}$ in LHMs are antiparallel to the
power flow ${\bf S}=\frac{1}{2} \text{Re}[{\bf E}\times{\bf
H}^\ast]$. Hence we conclude that the linear angular momentum flux
will reverse its direction in LHMs. Note that Eqs.~(\ref{momentum})
and (\ref{energy}) are valid only for lossless media, and its
application to lossy media produces unphysical phenomena such as a
negative energy in LHMs. In a lossy and dispersive LHM, the momentum
flow of a monochromatic wave is opposite to the power flow
direction. However, the momentum density may be parallel or
antiparallel to the power flow~\cite{Kemp2007}.

The cross product of this momentum density with the radius vector
${\bf r}$ yields an angular momentum flow. The angular momentum flow
in the $z$ direction depends upon the component of ${\bf
G}_\varphi$, such that
\begin{equation}
{\bf J}_z= {\bf r} \times {\bf T}_\varphi.
\end{equation}
Conservation of momentum at a material boundary ensures that the
tangential component of the wave momentum is
conserved~\cite{Kemp2005,Kemp2007}. Hence the angular momentum flow
still remains unchanged in the LHM. We can predict theoretically
that the orbital angular momentum for per phonon still remain $l
\hbar$. In order to accurately describe the angular momentum flow,
it is necessary to include material dispersion and losses. Thus a
certain dispersion relation, such as Lorentz medium model, should be
involved.

\begin{figure}
\includegraphics[width=8cm]{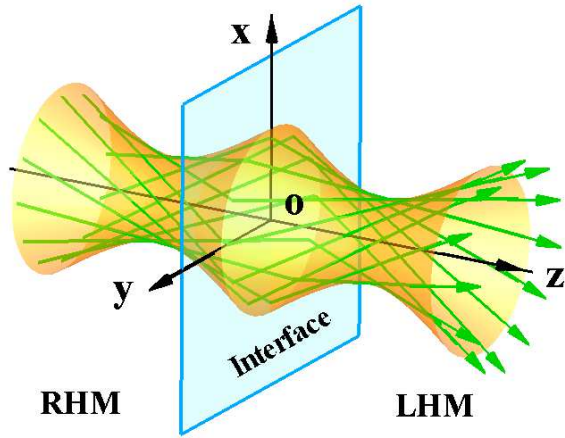}
\caption{\label{Fig2} (Color online) Ray optical model of
Laguerre-Gaussian beam in the RHM and the LHM. The rays (green
arrows) lies on a single-sheeted hyperboloid surface. Note that the
arrows indicate the direction of the Poynting vectors. When Rays
travel from a RHM to a LHM, the negative refraction results in the
reversed screw. The rays exhibit anti-clockwise screw in the RHM,
while he rays present clockwise screw in the LHM.}
\end{figure}

It has been shown that the trajectory at peak intensity becomes a
straight line skewed with respect to the beam axis in
RHMs~\cite{Courtial2000}. To obtain a better physical picture of the
straight trajectory in LHMs, the ray optical models of
Laguerre-Gaussian beam are plotted in Fig.~\ref{Fig2}. Within a ray
optical picture, the angular spectrum may be represented by skew
rays in the optical beam. The screwing behavior of rays can be
deduced from Eq.~(\ref{PVD}) in which we see that each ray having an
azimuthal angle $\theta=l/(n_{R,L} k_0 r)$ and a polar angel
$\eta=r/R$ with respect to the beam axis. Hence all rays lies on a
single-sheeted hyperboloid surface. Note that wave-vector and the
Poynting vector are parallel in the RHM and antiparallel in the LHM.
Energy conservation requires that the $z$ component of Poynting
vector must propagates away from the interface. Both wave-vector and
energy flow incident on a planar interface between a RHM and a LHM
will undergo negative refraction. Therefore all rays will reverse
its screwing fashion in the LHM (see Fig.~\ref{Fig2}).

\

To explore the reversed propagation dynamics in the LHM, we need to
move beyond ray tracing. The ray optical models neglect diffraction
and thus could not be used to predict precisely the spiral of
Poynting vector. To include diffraction, we had to use a more
accurate description of the electromagnetic field. It is well known
that the trajectory of the Poynting vector is described by the
spiral curve~\cite{Allen1992}. The relative value of the components
determines the trajectory of the Poynting vector. The spiral angle
of the Poynting vector is given by $\theta=\theta_0+ \kappa z$,
where $\kappa ={\partial\theta}/{\partial z}=S_\varphi/(r S_z)$ is
the rate of rotation. The period of the trajectory along the $z$
axis is $2\pi/ \kappa$. For a general Laguerre-Gaussian beam, the
rate of the azimuthal rotation is relate to the distance given by
\begin{eqnarray}
\frac{\partial\theta}{\partial z}=&&\frac{l}{n_{R,L} k_0
r^2}-\frac{\sigma |l|}{n_{R,L}k_0 r^2}+\frac{2 \sigma}{n_{R,L}k_0
w^2(z)}\nonumber\\
&& +\frac{4 \sigma}{n_{R,L}k_0
w^2(z)}\frac{L_{p-1}^{|l+1|}[2r^2/w^2(z)]}{L_p^{|l|}[2r^2/w^2(z)]}.\label{RAD}
\end{eqnarray}
Note that the first term is polarization independent. While the
last three terms depends on the polarization. For modes with
$p=0$, the final term is always zero.

\begin{figure}
\includegraphics[width=8cm]{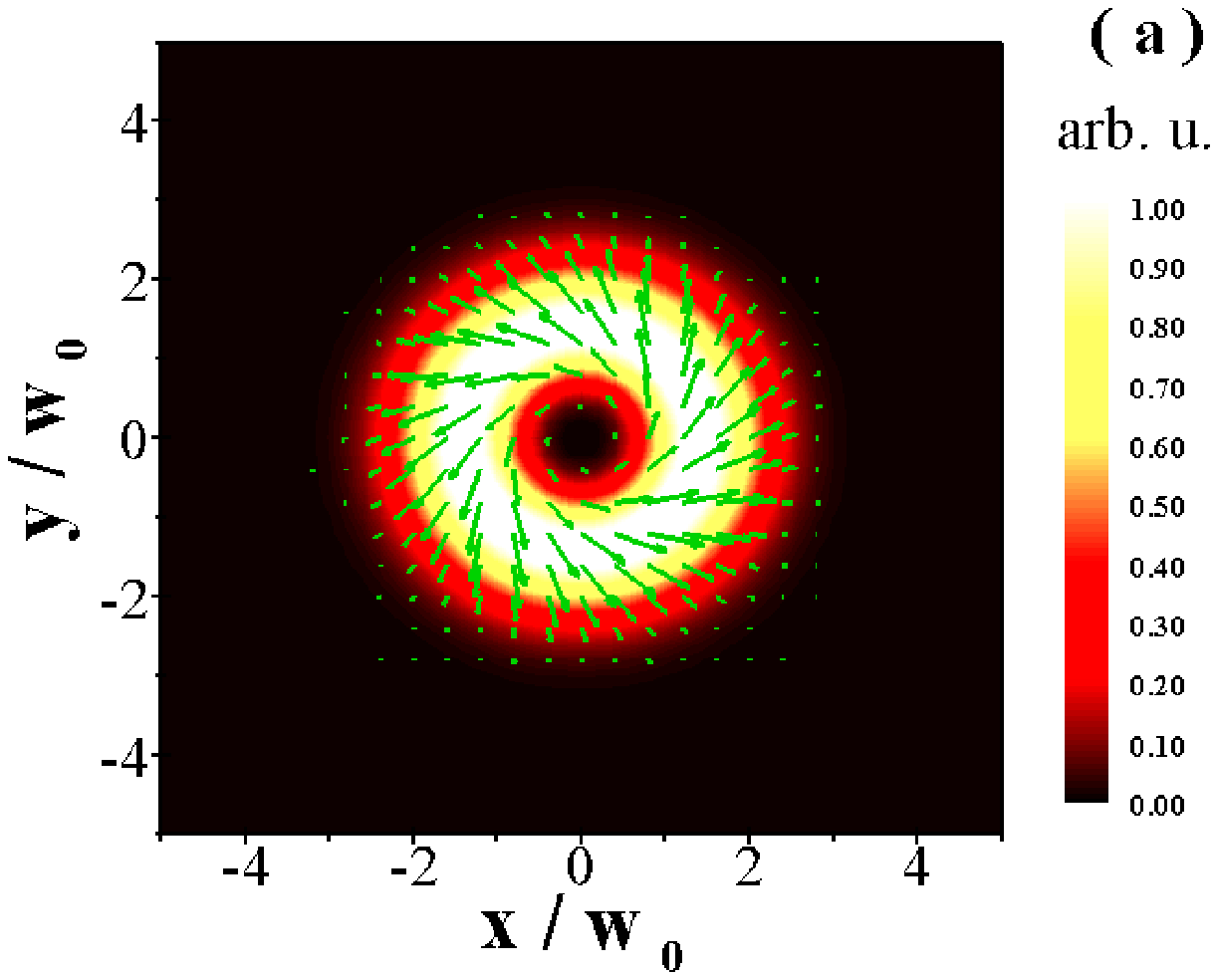}
\includegraphics[width=8cm]{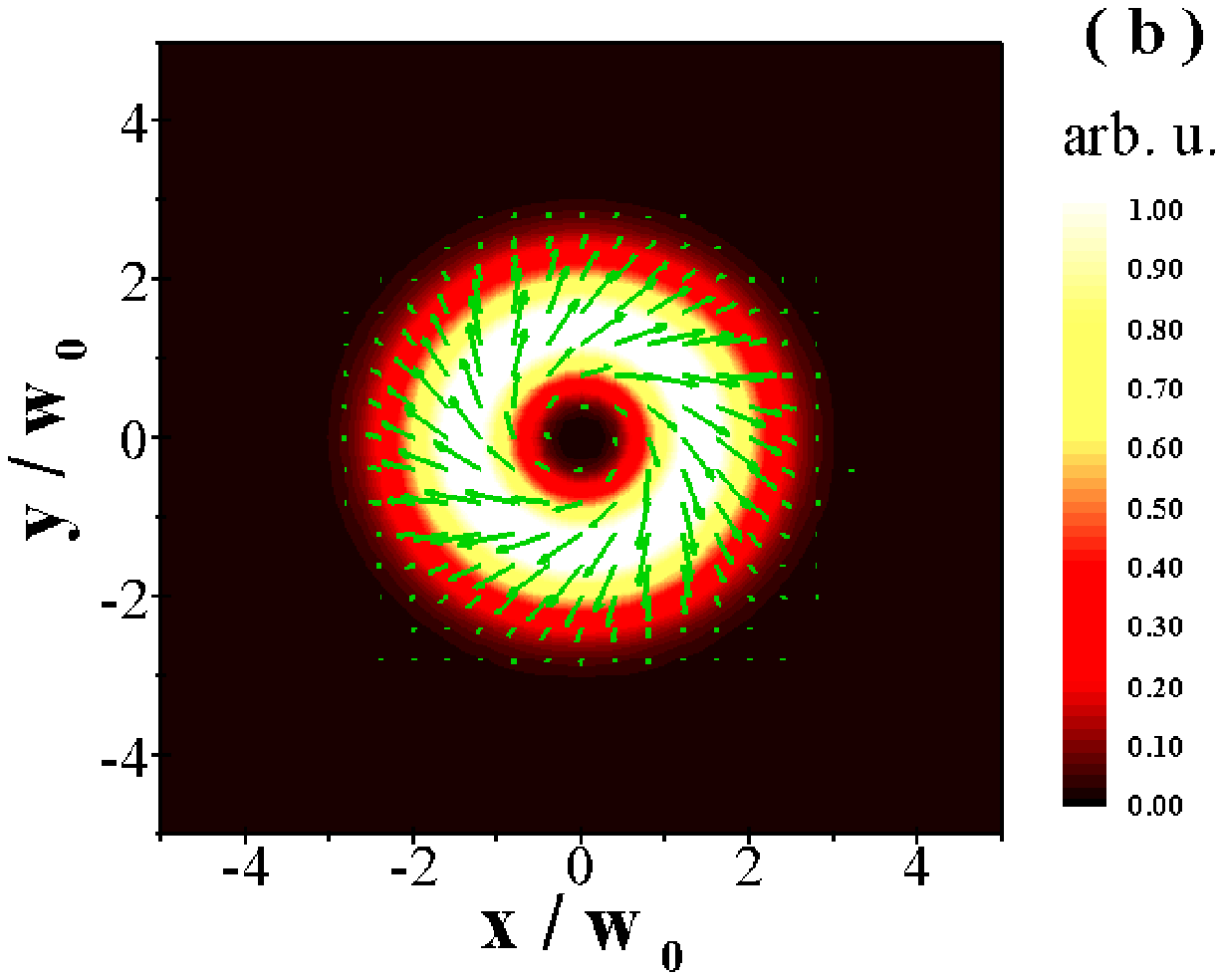}
\caption{\label{Fig3} (Color online) Numerically computed field
intensity distribution and transversal components of Poynting vector
(green arrows) for $l=3$ Laguerre-Gaussian beam. (a)The Poynting
vector exhibits anticlockwise spiral in RHM $z_1^{\ast}=n_R k_0
w_0^2 /2$. (b) The Poynting vector presents clockwise spiral in LHM
$z_2^{\ast}=|n_L| k_0 w_0^2 /2$. For the purpose of comparison, we
have chosen $n_L=-n_R$.}
\end{figure}

For a single-ringed Laguerre-Gaussian beam $p=0$ but $l\neq 0$, the
scaled radius of the peak intensity is given by
$r_{max}=\sqrt{|l|/2}w(z)$. We find that, for all values of $l$ and
$\sigma$, the rotation angle is given by
\begin{equation}
\theta_{max}=\frac{l}{|l|}\arctan\frac{z}{z_{R,L}}.\label{Gouy}
\end{equation}
Figure~\ref{Fig3} shows the vector fields illustrating the spiral
angle of the Poynting vector. The Poynting vector exhibits
anticlockwise spiral in the RHM as depicted in Fig.~\ref{Fig3}(a),
while presents clockwise spiral in the LHM as plotted in
Fig.~\ref{Fig3}(b). The theoretical analysis and numerical
calculations presented here coincide with experimental
observations~\cite{Leach2006}. For $l\neq 0$, Laguerre-Gaussian beam
have annular intensity profiles and as $r_{max}$ is typically much
greater than optical wavelength $\lambda$, the skew angle is
expected to be very small.

It can clearly be seen that the sense of spiral (clockwise or
anticlockwise) depends on the signs of $l$ and the Rayleigh length.
However, the amount of rotation of the poynting vector is
independent of the magnitude of $l$. It is interesting to note that
the Laguerre-Gaussian beam in LHMs will present the same fashion of
spiral as the counterpart with opposite topological charges in RHMs.
This is consist with the ray optical model describing the spiral of
the Poynting vector. Equation~(\ref{Gouy}) implies that the absolute
rotation for a single-ringed Laguerre-Gaussian beam at $z=z_{R,L}$
is $\pi/4$, regardless of $l$. When the far-field pattern of
Poynting vector is calculated it is found that the Poynting vector
is spiraled by $\pi/2$.

When $l=0$ and $p=0$, the Laguerre-Gaussian is identical to the
fundamental Gaussian beam, the spiral of the Poynting vector arises
from the effect of circular polarization~\cite{Allen2000}. The
azimuthal rotations in the RHM and the LHM are relate to the
distance can be obtained as $\theta=\sigma\arctan(z/z_{R})$ and
$\theta=\sigma\arctan(z/z_{L})$, respectively. It is intriguing to
note that the right-handed circularly polarized beam in LHMs will
present the same fashion of spiral as the left-handed circularly
polarized beam in RHMs. Hence we can describe quantitatively the
amount of spiral of the Poynting vector, from which we can determine
whether a material is a LHM or a RHM.

\begin{figure}
\includegraphics[width=8cm]{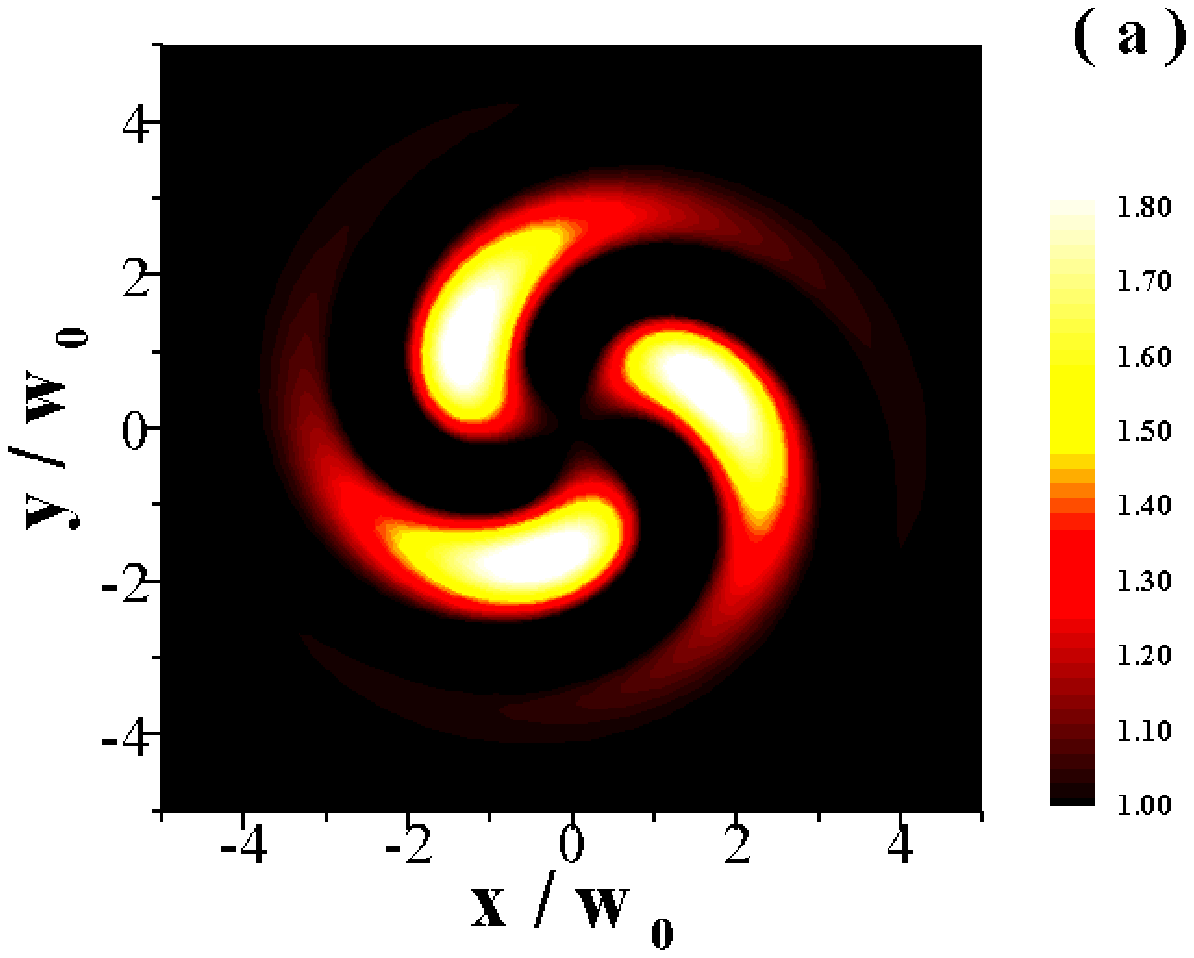}
\includegraphics[width=8cm]{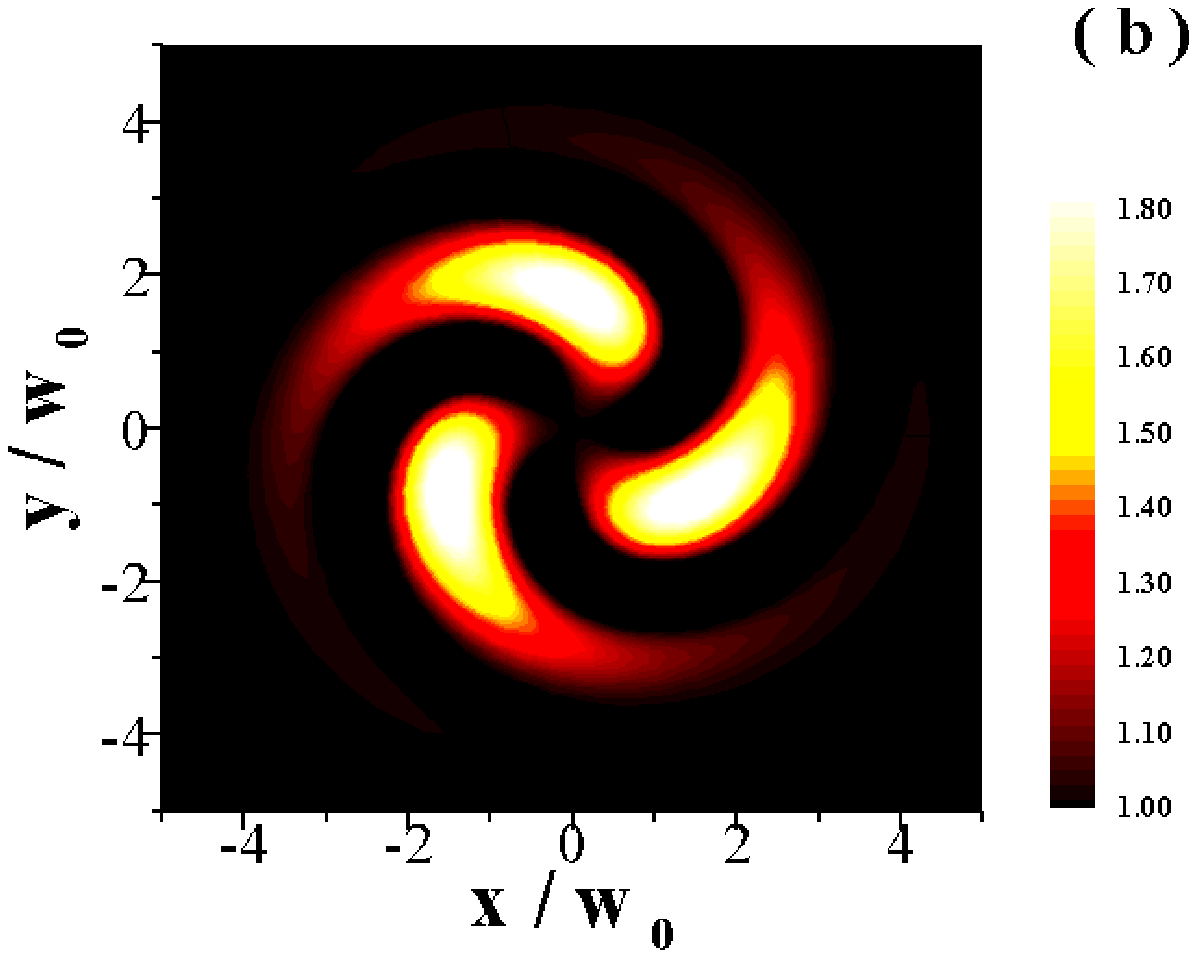}
\caption{\label{Fig4} (Color online) Interfering a Gaussian beam and
a Laguerre-Gaussian beam of azimuthal index $l=3$ will produce the
vortex field with three spiral arms. (a) The field distribution at
plane of $z_1^{\ast}=n_R k_0 w_0^2 /2$. (b) The field distribution
at plane of $z_2^{\ast}=|n_L| k_0 w_0^2 /2$. When the vortex field
enters the LHM, the rotation changes its fashion.}
\end{figure}

Interfering the Laguerre-Gaussian beam with a fundamental Gaussian
beam will transform the azimuthal phase variation of the pattern
into an azimuthal intensity variation. Hence the helical phase
ultimately results in an vortex field with $|l|$ spiral
arms~\cite{Padgett1996,Soskin1997,Macdonald2002}. The intriguing
properties strongly motivate us to explore the vortex field
propagation in LHMs. A simulation of the vortex field rotation
produced by interfering a fundamental Gaussian beam and an
Laguerre-Gaussian beam of azimuthal index $l=3$ is shown in
Fig.~\ref{Fig4}. It can be seen that the vortex has three spiral
arms, which is a result of the mismatch between the wave-fronts of
the Laguerre-Gaussian beam and the Gaussian beam. The reversed
screwing wave-fronts will directly cause an inverse rotation of the
vortex filed in the LHM. The vortex will always has a spiral shape
unless the wavefronts of the two beams have the same curvature. For
example, the vortex field at a focusing waist will exhibit $|l|$
intense spots. After the vortex propagating through the focusing
waist, the spiral arms will change their shapes.

Now let us consider how to modulate the rotation of the vortex
field. As we change the path length of the Gaussian beam, the spiral
arms will rotate around the propagating axis. This is analogous to
altering the phase different between the Laguerre-Gaussian beam and
the Gaussian beam~\cite{Macdonald2002}. The spiral arms repeat every
$\lambda/|n_L|$ in the LHM, but only rotate fully after propagating
$|l|\lambda/|n_L|$. A path length change in the Gaussian beam of $3
\lambda/|n_L|$ will cause the pattern to rotate through $2 \pi$ and
$-2 \pi$ in RHM and LHM, respectively. The vortex presents an
anticlockwise rotation in the RHM, while exhibits a clockwise
rotation in the LHM. Once the vortex field enter the LHM, it will
reverse its rotation fashion.

\section{Conclusions}
In conclusion, we have investigated the reversed propagation
dynamics of Laguerre-Gaussian beam in LHMs. We have introduced the
concepts of negative Gouy-phase shift to describe the propagation of
Laguerre-Gaussian beam in LHMs. The negative phase velocity and
negative Gouy-phase shift caused inverse screw of wave-fronts,
reversed spiral of Poynting vector, and inverse rotation of vortex
field. At a RHM-LHM interface, direct calculation of Maxwell's
equations dictates the wave-vector and energy flow undergoes
negative refraction. Consequently, inside the LHMs, the screw of
wave-fronts, the spiral of Poynting vector, and the rotation of
vortex will reverse their direction. We have shown that the Poynting
vector of Laguerre-Gaussian beam in LHMs will present the same
fashion of spiral as the counterpart with opposite topological
charges in RHMs. Conservation of momentum at the boundary ensure
that the tangential component of the wave momentum is conserved. We
have found that although the linear momentum reverses its direction,
the angular momentum still remains unchanged. Since the photons in
Laguerre-Gaussian beam possess angular momentum, the reversed
propagation dynamics may offer new fundamental insights into the
nature of LHMs.

\begin{acknowledgements}
The authors are sincerely grateful to Professors Wei Hu and Zhenlin
Wang for many fruitful discussions. This work was supported by
projects of the National Natural Science Foundation of China (Grants
Nos. 10535010, 10576012, 10674045, 10775068, and 60538010), the 973
National Major State Basic Research and Development of China (Grant
No. G2000077400), and Major State Basic Research Developing Program
(Grant No. 2007CB815000).
\end{acknowledgements}

\end{document}